\documentclass[letter]{aa}  

\usepackage{graphicx}
\usepackage{xcolor}
\usepackage{float}
\usepackage{rotating}
\usepackage{booktabs}
\usepackage{mathtools}
\usepackage{placeins}
\usepackage{bm}
\usepackage{longtable}
\usepackage{adjustbox}
\usepackage{supertabular}
\usepackage{multirow}
\usepackage{tabularx}
\usepackage{url}
\usepackage{esdiff}
\usepackage{amsmath}
\usepackage{siunitx}
\usepackage[switch]{lineno}

\usepackage{hyperref}
\hypersetup{
  colorlinks=true,
    linkcolor=blue,
    citecolor=blue,
    urlcolor=magenta}

\usepackage{natbib}
\usepackage{txfonts}

\defcitealias{Tokuno2022}{TT22}

\begin{document} 

   \title{Core-envelope coupling of gravito-inertial waves \\ in pre-main-sequence solar-type stars}
    
   \titlerunning{}

   \author{S.N.~Breton\inst{1}
          \and 
          C.~Pezzotti\inst{2}
          \and
          S.~Mathis\inst{3}
          \and
          L.~Bugnet\inst{4}
          \and
          M.P.~Di Mauro\inst{5}
          \and
          J.~Joergensen\inst{6}
          \and 
          K.~Zwintz\inst{6}
          \and
          A.F.~Lanza\inst{1}
          }
    \institute{INAF – Osservatorio Astrofisico di Catania, Via S. Sofia, 78, 95123 Catania, Italy \\
    \email{sylvain.breton@inaf.it} 
    \and
    STAR Institute, Université de Liège, Liège, Belgium
    \and
    Universit\'e Paris-Saclay, Universit\'e Paris Cit\'e, CEA, CNRS, AIM, 91191, Gif-sur-Yvette, France
    \and 
    Institute of Science and Technology Austria (ISTA), Am Campus 1, Klosterneuburg, Austria
    \and 
    INAF – IAPS, Istituto di Astrofisica e Planetologia Spaziali, Via del Fosso del Cavaliere 100, 00133 Roma, Italy
    \and 
    Universität Innsbruck, Institut für Astro- und Teilchenphysik, Technikerstraße 25, 6020 Innsbruck, Austria
    }

   \date{}

 \abstract{
    After the recent detection of solar equatorial Rossby waves, a renewed interest has been brought to the study of gravito-inertial waves propagating in the convective envelope of solar-type stars. In particular, the ability that some of these envelope gravito-inertial modes have to couple with the ones trapped in the radiative interior might open new windows to probe the deep-layer dynamics of solar-type stars. The possibility for such a coupling to occur is particularly favoured in pre-main sequence (PMS) solar-type stars. Indeed, due to the contraction of the protostellar object, they are able to reach large rotation frequencies before nuclear reactions are ignited and magnetic braking becomes the driving mechanism for their rotational evolution. In this work, we therefore study the coupling between the envelope inertial waves and the radiative interior $g$ modes in PMS stars, focusing on the case of prograde dipolar modes. We consider the case of 0.5~$\rm M_\odot$ and 1~$\rm M_\odot$ PMS models, each with three different scenarios of rotational evolution. We show that, for stars that have formed with a sufficient amount of angular momentum, this coupling can occur in frequency ranges that are accessible to space-borne photometry, creating inertial dips in the period spacing pattern. With an asymptotic analysis we characterise the shape of these inertial dips to show that they depend on rotation and on the stiffness of the convective-radiative interface.
 
 }

 \keywords{asteroseismology -- Stars: rotation -- Stars: oscillations (including pulsations)}

   \maketitle

\section{Introduction \label{sec:introduction}}

The recent detection of equatorial Rossby waves \citep[e.g.][]{Papaloizou1978,Saio1982} in the Sun \citep{Loptien2018} has insufflated a reinvigorated interest in the study of gravito-inertial modes of oscillation in solar-type stars \citep[e.g.][]{Jain2024}. In particular, \citet{Hindman2023} demonstrated that thermal Rossby waves, a class of gravito-inertial waves trapped in the convective envelope \citep{Hindman2022,Hanasoge2026}, distinct from equatorial Rossby waves and also referred as convective modes, are able to couple with the prograde $g$ modes trapped in the radiative interior. The observation and characterisation of such a coupling between radiative interior $g$ modes and envelope inertial modes of oscillation would open new perspectives to probe the deepest stellar layers. However, in a star rotating as slowly as the Sun, this coupling can only act at very low frequencies, making a detection attempt observationally challenging in the context of asteroseismology, because it would require extremely long time series, as well as exceptional instrumental stability at low frequency. On the contrary, young solar-type stars are rotating significantly faster before they enter the main sequence \citep[e.g.][]{Gallet2015} and spin down under the action of magnetic braking \citep[e.g.][]{Weber1967,Skumanich1972}. This would enable the eigenfrequencies of the inertial modes to extend towards larger values, in frequency intervals more easily accessible to space-borne photometry \citep[see e.g.][for a candidate signature in a M dwarf]{Messina2024}. With respect to other stages of evolution, a relatively small number of studies has been dedicated to the asteroseismology of pre-main sequence (PMS) stars \citep[see][for a review]{Zwintz2022}. Progenitors of the systems we observe on the main sequence, they are nevertheless of primary importance to understand the history of older targets. 
%This emphasise the importance to improve our theoretical understanding of the oscillation mode signatures in such objects in order to fruitfully identify them in the data. 

In this letter, we show that, in PMS stars, $g$ modes and envelope inertial waves are able to couple in frequency ranges that can reasonably be accessed by photometric instruments such as \textit{Kepler}/K2 \citep{Borucki2010,Howell2014}, the Transiting Exoplanet Survey Satellite \citep[TESS,][]{Ricker2015}, or the PLanetary Transits and Oscillations of star mission \citep[PLATO,][]{Rauer2025}. In Sect.~\ref{sec:stellar_model}, we present the stellar models we use for the study, while in Sect.~\ref{sec:mode_coupling}, we compute the eigenfunctions of the gravito-inertial modes and we show that the radiative-convective coupling induces an inertial dip that can be described asymptotically by a Lorentzian profile depending on the rotation frequency and on the stiffness of the convective-radiative interface. Conclusion and perspectives are drawn in Sect.~\ref{sec:conclusion}. 

\section{Stellar models \label{sec:stellar_model}}

\subsection{Structure models}

To illustrate the possibility of a coupling between $g$ modes and envelope inertial waves, we consider the case of a K-type and a G-type PMS progenitors at solar metallicity, with mass $M_\star = 0.5 \, \mathrm{M}_\odot$ and $1 \, \mathrm{M}_\odot$, respectively. Their properties are summarised in Table~\ref{tab:properties}. We use the PMS evolutionary tracks computed by \citet{Steindl2021} with the Module for Experiments in Stellar Astrophysics \citep[MESA,][]{Paxton2011}. These PMS models assume constant accretion for the star to reach its actual mass, starting from an initial seed with mass 0.01~$\rm M_\odot$. The models are then evolved until the start of the hydrogen burning phase at the zero-age main sequence (ZAMS). The evolutionary tracks for the two selected models are represented in Fig.~\ref{fig:kiel_diagram}, from the end of the accreting phase to the ZAMS. We consider the structure of the 0.5~$\rm M_\odot$ model at 21.4~Myr and of the 1~$\rm M_\odot$ one at 12.4~Myr. We note that, in this work, we limit ourselves to the case of structure models with a radiative interior and a convective envelope, the complex mode coupling that would be induced by the combination of a convective core, a radiative interior and a convective envelope being out of the scope of this work.

\subsection{Rotational evolution}

\begin{figure}[ht!]
    \centering
    \includegraphics[width=0.98\linewidth]{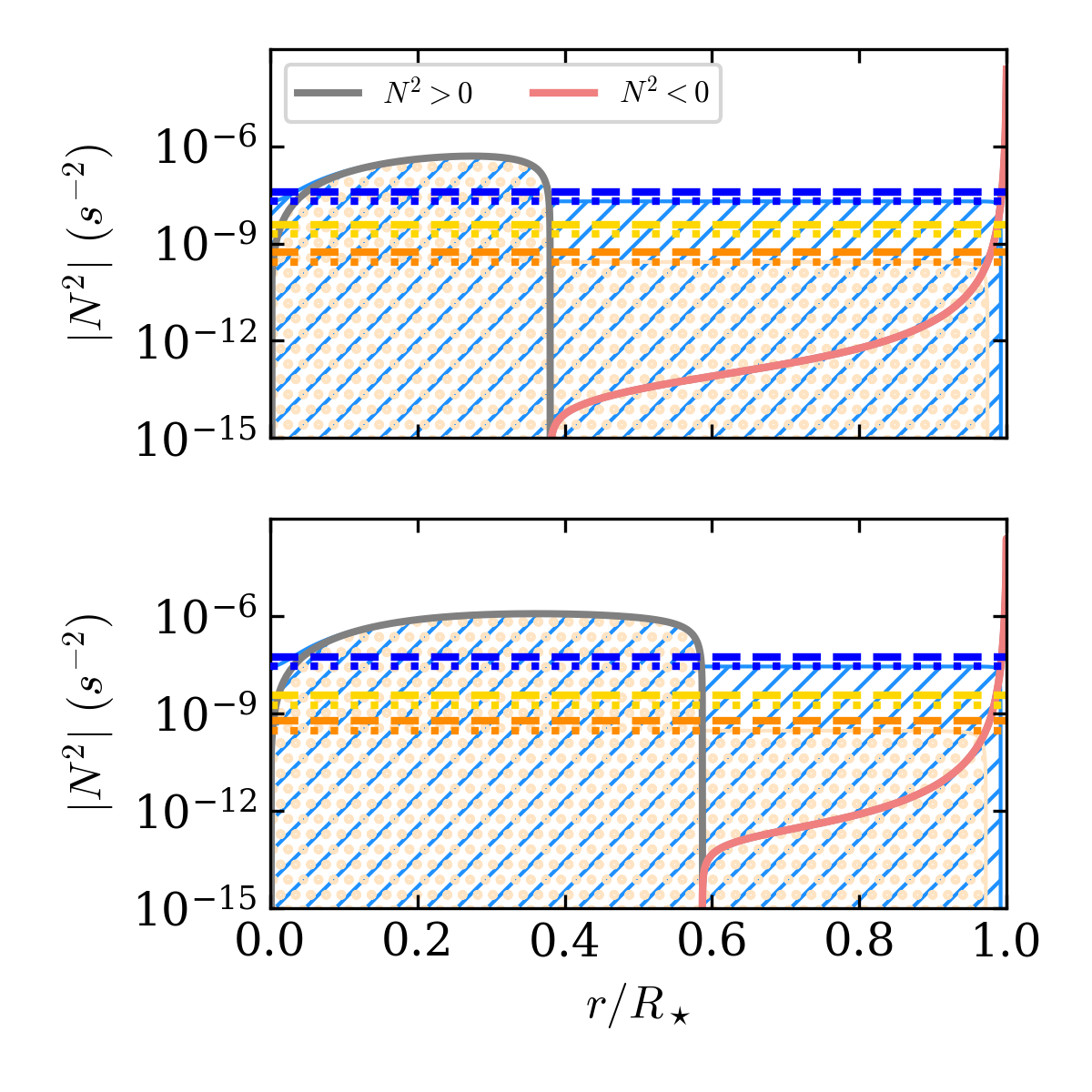}
    \caption{Modulus of the squared Brunt-Väisälä frequency, $|N^2|$ for the 0.5~$\rm M_\odot$ model (top) and the 1~$\rm M_\odot$ model (bottom), with regions where $N^2 \geq 0$ in gray and regions where $N^2<0$ in red. The squared Coriolis frequency, $4\Omega^2$, is indicated by the horizontal dashed line for the slow (orange), {the intermediate (yellow)}, and the fast (blue) case, while the horizontal dotted lines corresponds to $2\Omega^2$. The propagation region of gravito-inertial waves as a function of frequency is highlighted by the hatched areas (orange dots and blue hatches for the slow and fast cases, respectively).}
    \label{fig:structure}
\end{figure}

The rotational evolution of the two models is computed a posteriori, considering that the radiative-convective coupling timescale for transport of angular momentum is short enough to assume solid-body rotation \citep[e.g.][]{Spada2011}. We account for wind braking and for a disc-coupling phase at the end of the accretion phase. We consider the cases of a {slow, an intermediate, and a fast initial rotation rate, with $\Omega_{\rm \star, init} = 3.2$, $8$, and  $18 \, \Omega_\odot$, respectively}. This rotation rate is set to be constant during the disc-coupling phase, which lasts 6~Myr in the slow and intermediate cases and 2~Myr in the fast case. Then, during the PMS phase, the star spins up due to contraction, although a fraction of its angular momentum is already lost due to stellar magnetised wind. The prescription used to account for magnetic braking of the stellar surface is the one by \citet{Matt2015, Matt2019}, whose free parameters have been calibrated to reproduce the surface rotation rate of the Sun. The $\rm \Omega_{\star, init}$ values and the relative disc-locking timescales for the fast and slow rotators have been calibrated to reproduce the spread of surface rotation rates observed for stars in open clusters and stellar associations \citep{Gallet2015, Eggenberger2019a}.

The rotational evolution during the span of the evolutionary tracks represented in Fig.~\ref{fig:kiel_diagram} is illustrated in Fig.~\ref{fig:rotational_evolution}. For the reference age and structure we consider, the models with slow initial rotation rate have a rotation frequency $\Omega_\star = 4.5$ and $4.7 \, \Omega_\odot$, {the models with intermediate rotation have $\Omega_\star = 12.0$ and $11.6 \, \Omega_\odot$} and the models with fast rotation have $\Omega_\star = 38.4$ and $45.4 \, \Omega_\odot$,  for the 0.5~$\rm M_\odot$ and 1~$\rm M_\odot$ models, respectively. These last values corresponds to ratios $\Omega_\star / \Omega_{\rm c} = 0.11$ and $\Omega_\star / \Omega_{\rm c} = 0.22$, where $\Omega_{\rm c} = (GM_\star / R_\star^3)^{1/2}$ is the critical Keplerian frequency. Although this places the fast case at the limit of the $\Omega_\star / \Omega_{\rm c} \ll 1$ assumption, in what follows, we neglect the impact of the centrifugal acceleration in order to model our star using 1D non-deformed profiles and therefore to preserve the simplicity of the physical picture we aim at discussing here. 

\section{Mode coupling \label{sec:mode_coupling}}

In what follows, we use the equatorial model presented in Appendix~\ref{appendix:system_derivation} \citep[see also][Mathis et al. submitted]{Ando1985,Jain2024}. The problem is restricted to the equatorial frame in order to account for the full Coriolis force \citep[][Mathis et al. submitted]{Ando1985}. This allows us to grasp qualitatively the physical picture of the impact that the non-traditional component of the Coriolis force has on {this waves, that are confined close to the equator \citep{Jain2023}}. We numerically solve the linearised Euler equations in the \citet{Cowling1941} approximation, assuming spherical symmetry for the equilibrium quantities. 
\begin{align}
    \label{eq:xi_r}
    \diff{\xi_r}{r} &= - \left( \frac{1}{r} \left( 2 + \frac{m \zeta}{\omega} \right) + \frac{1}{\Gamma_1} \diff{\ln p}{r} \right) \xi_r + \frac{1}{\rho c_s^2} \left ( \frac{S_{\tilde{\ell}}^2}{\omega^2} - 1 \right) p' \; , \\
    \label{eq:p_prime}
    \diff{p'}{r} &= \rho \Bigg(\omega^2 - N^2 - 2\Omega \zeta \Bigg) \xi_r + \left( \frac{1}{\Gamma_1} \diff{\ln p}{r} + \frac{2 \Omega m}{\omega r} \right) p' \; ,
\end{align}
where $\omega$ is the wave local angular frequency, $r$ is the stellar radius, $p$ the equilibrium pressure, $\rho$ the equilibrium density, $c_s$ the sound speed, $N^2$ the Brunt-Väisälä frequency, $\Gamma_1 = (\partial \ln p / \partial \ln \rho)_{\rm ad}$ the adiabatic exponent, $\Omega$ the rotation profile and $\zeta = 2\Omega + r (\mathrm{d}\Omega / \mathrm{d}r)$ the stellar vorticity. The local frequency $\omega$ is related to the inertial frequency, $\omega_{\rm in}$, through the relation  $\omega = \omega_{\rm in} - m \Omega$. The perturbed quantities for which eigenfunctions are searched for are $\xi_r$, the radial displacement, and $p'$, the Eulerian pressure perturbation. Finally, $m$ is the azimuthal number and ${\tilde{\ell}}$ is the global horizontal degree, the Lamb frequency $S_{\tilde{\ell}}$ therefore writes $S_{\tilde{\ell}}^2 = {\tilde{\ell}} ({\tilde{\ell}} + 1) c_s^2 / r^2$ with the sound speed $c_s = \sqrt{\Gamma_1 p / \rho}$. In the case without rotation, Eq.~(\ref{eq:xi_r}) and (\ref{eq:p_prime}) simply correspond to the second-order system of stellar oscillations in the Cowling approximation \citep[e.g.][]{Aerts2010}. In what follows, we consider dipolar prograde modes with ${\tilde{\ell}} = m = 1$.

\begin{figure}[ht!]
    \centering
    \includegraphics[width=0.98\linewidth]{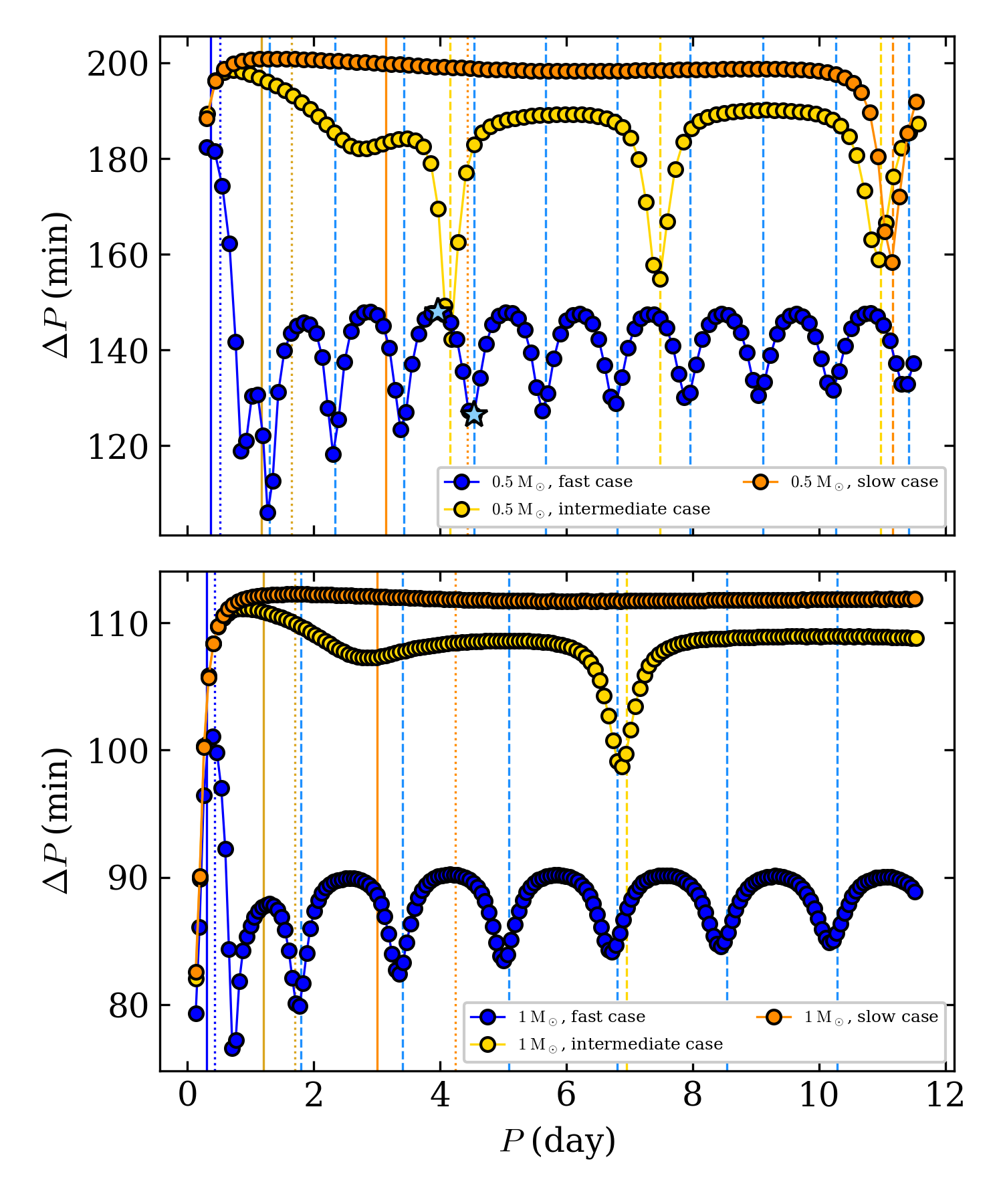}
    \caption{\textit{Top:} $\Delta P$ vs $P$ diagram for the ${\tilde{\ell}} = m = 1$ modes in the 0.5~$\rm M_\odot$ model, in the co-rotating frame. The dots correspond to the modes computed considering the full extent of the stellar structure, with the slow case in orange, {the intermediate case in yellow}, and the fast case in blue. The vertical dashed lines mark the period of the modes computed considering only the convective envelope. The position of the Coriolis frequency, $2 \Omega$, is {highlighted for the slow, the intermediate and the fast cases (vertical thick lines in orange, yellow, and blue, respectively)} while the $\sqrt{2}\Omega$ frequency is indicated with the vertical dotted lines. 
    \textit{Bottom:} Same as top panel for the $1~\mathrm{M}_\odot$ model. In the top panel, the location of the modes with the eigenfunctions represented in Fig.~\ref{fig:eigenfunction_coupling_example} are highlighted with the stellar symbol in light blue.
    }
    \label{fig:modes_both_profiles}
\end{figure}

\begin{figure}[ht!]
    \centering
    \includegraphics[width=0.98\linewidth]{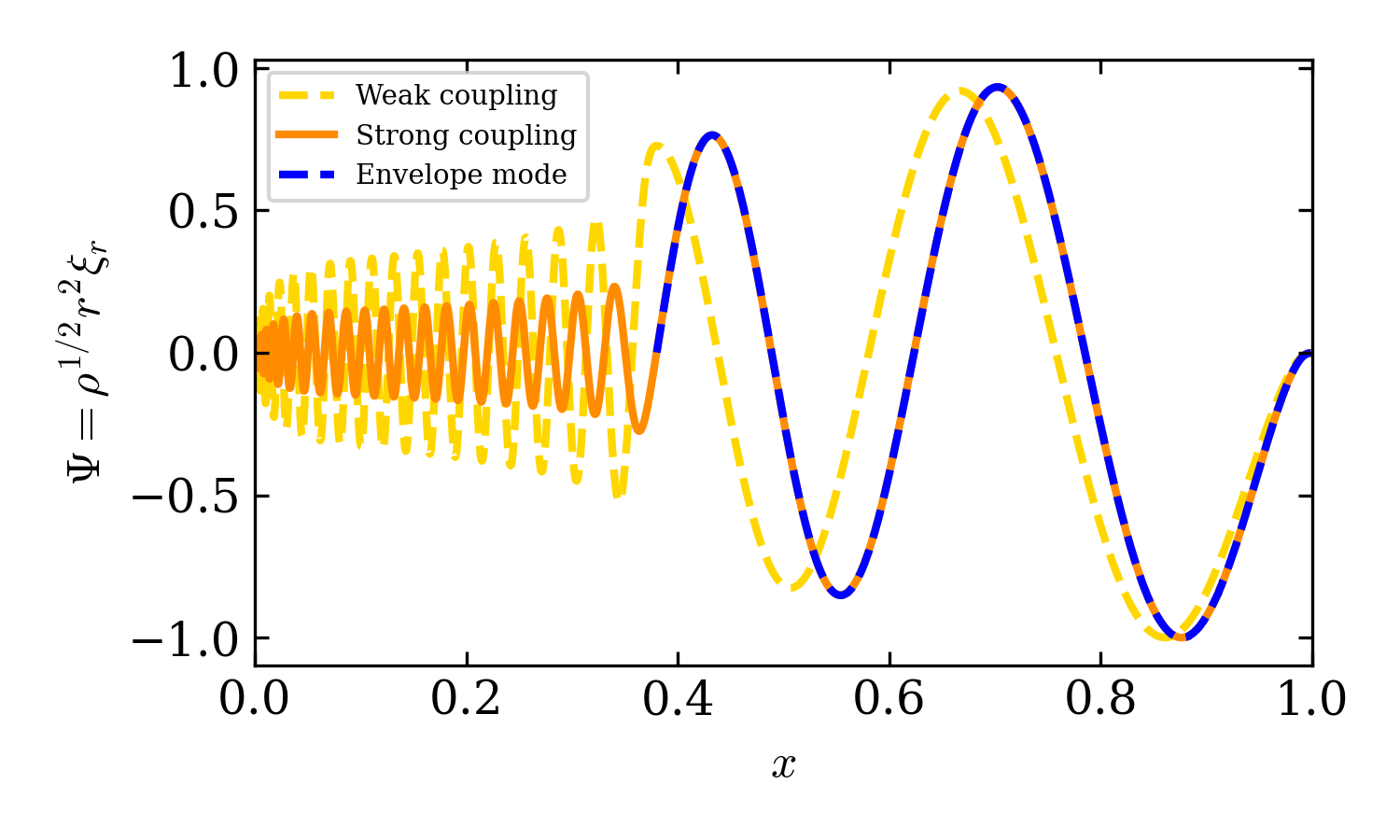}
    \caption{Comparison between the eigenfunctions of the $n = - 46$ (orange) and the $n = -40$ (dotted grey) modes. The eigenfunction of the $n = -4$ envelope mode is shown in dashed blue.}
    \label{fig:eigenfunction_coupling_example}
\end{figure}

As discussed in Appendix~\ref{appendix:propagation}, the propagation of the gravito-inertial waves will be constrained by the shape of the buoyant and Coriolis cavities, related to the $N$ and $2\Omega$ frequencies, respectively. Roughly, from Eq.~(\ref{eq:propagation_condition_bulk}), for dipolar prograde modes, the mode will have an oscillating behaviour when 
\begin{equation}
    \omega^2 <  N^2 + 4\Omega^2 \left(1 - \frac{m^2}{{\tilde{\ell}} ({\tilde{\ell}}+1)}\right) = N^2 + 2\Omega^2 \; .
\end{equation}
In Fig.~\ref{fig:structure}, we show the propagation regions of the waves as a function of frequency. We note that, in the bulk of the convective envelope, $|N^2| \ll 4\Omega^2$, where $4\Omega^2$ is the squared Coriolis frequency. In the radiative interior, $N^2 \gg 4 \Omega^2$ in the slow case but not in the fast case. In addition, below a given radius, we have $N^2 < 4 \Omega^2$. It is also visible that the convective envelope of the 0.5~$\rm M_\odot$ model has a larger extension, with its bottom located at $r_{\rm cz} = 0.38 \, R_\star$ while $r_{\rm cz} = 0.59 \, R_\star$ for the 1~$\rm M_\odot$ stellar model.

The method we use to numerically compute the eigenfunctions of our stellar models is detailed in Appendix~\ref{appendix:numerical_approach}. For each stellar model, we first search for eigenfunctions that are solutions to the system defined by Eqs.~(\ref{eq:xi_r}) and (\ref{eq:p_prime}) considering the full extent of the stellar structure, and we then restrict the domain to the extent of the convective envelope. The eigenfunctions are computed with the open source \texttt{mascaret}\footnote{The code source is accessible at \url{https://gitlab.com/sybreton/mascaret} and the documentation at \url{https://mascaret.readthedocs.io}.} module which implements a double shooting scheme to identify the eigenvalues of the system \citep{Press1992}. At low frequency, we limit our search to $\omega/(2 \pi) = 1 \; \mu$Hz, which corresponds to a period in the co-rotating frame of about 11.57~days. Given that $m=1$, in the inertial frame, the eigenfrequencies will be shifted towards larger values. We compute $\Delta P$ as the period spacing between the periods in the co-rotating frame, $P$, of modes with consecutive radial order $n$. For {the three rotation regimes}, the $\Delta P$ versus $P$ diagram we obtain for the computation considering the full extent of the stellar structure is represented in Fig.~\ref{fig:modes_both_profiles}. For comparison, we also show in Fig.~\ref{fig:delta_p_inertial_frame_frequency_xaxis}, the inertial period spacing, $\Delta P_{\rm in}$, versus $\omega_{\rm in}$ diagram. $\Delta P_{\rm in}$ will be the directly observable quantity. 
In the fast {and intermediate cases}, the $\Delta P$ pattern is strongly affected by the coupling with the envelope inertial modes. The periods of the eigenmodes obtained when restricting the computation to the convective envelope are highlighted in Fig.~\ref{fig:modes_both_profiles}. Around the eigenfrequencies of these envelope eigenmodes, the coupling between the $g$ modes trapped in the radiative interior and a gravito-inertial mode from the envelope are clearly visible as sudden drops in the $\Delta P$ pattern, referred as inertial dips \citep{Ouazzani2020,Saio2021}. Away from the dips, the $\Delta P$ pattern also significantly differs between the slow and the fast cases, reflecting the contribution from the Coriolis acceleration to the mode dynamics: in the radiative interior, the Coriolis contribution enforces the effective stratification of the medium, reducing the period spacing between consecutive modes (see Eq.~(\ref{eq:effective_stratification})). In addition, notwithstanding the strength of the coupling, all modes with $P > 1 / (2\Omega_\star)$ are in the subinertial regime and the corresponding waves remain propagative even in the convective envelope. Concerning the slow rotation case, there is no inertial dip at periods shorter than 12 days for the 1~$\rm M_\odot$ model. Nevertheless, the first inertial dip is visible around $P =  11$~days for the 0.5~$\rm M_\odot$ model. It is possible to show that each of these inertial dips can be described asymptotically with a Lorentzian profile \citep[][see Appendix~\ref{appendix:inertial_dips} for the derivation in the case of our equatorial formalism]{Tokuno2022}
\begin{equation}
    \Delta P = \Delta P_0 \left( 1 - \frac{\Delta P_0 \sigma / \pi}{(P - P_* + \sigma / \sqrt{3})^2 + \sigma^2 + \Delta P_0 \sigma / \pi} \right) \; ,
\end{equation}
where $\Delta P_0$ is given by Eq.~(\ref{eq:tassoul_delta_p}), and the parameter $\sigma$ depends on $\Omega$ and on the stiffness of the squared Brunt-Väisälä frequency, $\mathrm{d}N^2 / \mathrm{d}r$, at the convective-radiative interface, as shown by Eq.~(\ref{eq:epsilon}) and (\ref{eq:sigma}). As illustrated in Fig.~\ref{fig:dips_both_models}, this asymptotic approximation describes fairly well the shape of the inertial dips we obtained from the numerical computations.

Finally, we note the existence of a first dip for both structure models, in the fast case, that is not connected to the eigenfrequency of the envelope modes. We interpret this by recalling that, in the central regions of the star, the shape of the resonant cavity is strongly modified by the Coriolis force with respect to the $\Omega = 0$ case (see Fig.~\ref{fig:structure}). This first dip therefore most likely reflects the sensitivity of the mode to this deformation.  

In Fig.~\ref{fig:eigenfunction_coupling_example}, we finally compare two eigenfunctions of the 0.5~$\rm M_\odot$ model. The location of the two modes in the $\Delta P$ vs $P$ diagram was highlighted in Fig.~\ref{fig:modes_both_profiles}. The first one, with radial order $n = - 46$ is strongly coupled with the envelope mode with $n = -4$, while the second one, with $n = - 40$, is weakly coupled. The oscillating character of $\xi_r$ in the convective envelope is clearly visible for both the strongly and weakly coupled modes, with several radial nodes located above $r_{\rm cz} = 0.38 \, R_\star$, highlighting the fact that both modes are in the subinertial regime. We note that the eigenfunction of the strongly coupled mode closely follow the one of the envelope mode in the convective zone.
%Nevertheless, the relative displacement $\xi_r$ reaches a larger amplitude for the strongly coupled mode. 
%We finally note that the $g$ mode component of the coupled mode is high order, allowing for asymptotic description, while this is not the case for the envelope inertial component (see also Fig.~\ref{fig:delta_p_envelope_modes}).

\section{Discussion and conclusion \label{sec:conclusion}}

The coupling that is observed here between the radiative interior and the convective envelope of the PMS stars appears analogous to the coupling between the convective core and radiative envelope that was evidenced in $\gamma$~Dor stars \citep[e.g.][]{Ouazzani2020,Saio2021,Tokuno2022,Galoy2024,Barrault2025}. Different scenarios lead to different observable consequences for the mode amplitudes. On the one hand, convection might transfer energy to the buoyant cavity through penetrative convection \citep[e.g.][]{Alvan2014,Pincon2016,Augustson2020,Breton2022simuFstars} or bulk turbulence \citep[e.g.][]{Belkacem2009,Lecoanet2013,Mathis2014,Augustson2020}, as in the traditional $g$ mode picture. On the other hand, convection might be able to transfer more efficiently its energy to the envelope inertial modes \citep{Philidet2023,Fuentes2025}. Therefore, if the excitation in the $g$ mode resonant cavity is the dominating process, the modes should be able to emerge with similar surface amplitudes, even for the ones with a weak coupling. On the contrary, if the excitation energy is channelled first to the envelope inertial waves before propagating to the radiative interior, the modes with a strong coupling should have larger surface amplitude. Given that PMS solar-type stars are magnetically active, future works should be aimed at evaluating the impact that an internal magnetic field will have on the shape of the dip \citep[e.g.][]{Barrault2025b}.

In this work, we computed the eigenfrequencies for gravito-inertial dipolar prograde modes in 0.5~$\rm M_\odot$ and 1~$\rm M_\odot$ PMS stars in order to illustrate how the coupling between the eigenmodes of the radiative interior and the ones of the convective envelope of such objects will be affected by rotation. 
Although the fast rotation regime introduces centrifugal effects that typically deform the stellar cavity, we have neglected them to isolate the Coriolis-driven dynamics. The centrifugal distortion would introduce a systematic shift in the eigenfrequencies and slightly modify the cavity geometry, but the fundamental mechanism of the core-envelope coupling and the resulting period-spacing pattern would persist.
If the modes are sufficiently excited in order to induce detectable brightness variations, for fast PMS rotators, the coupling will occur in a frequency range that is accessible to space-borne photometry. 
%The signature we evidenced here should be searched for in young open clusters and star forming regions. 
The imprint it leaves on the mode pattern should be able to provide important information on the mode excitation mechanism. In addition, the analysis we carried out in this work demonstrates that these coupled modes should represent powerful seismic probes. Indeed, while the shape of the inertial dips is directly related to the stiffness of the convective-radiative interface, it will also be possible to infer $\Delta P_0$ from the shape of the Lorentzian profile in the $\Delta P$ pattern, this in order to probe the characteristics of the deep radiative interior.  

\begin{acknowledgements}
The authors want to thank the anonymous referee for useful comments. SNB acknowledges support from PLATO ASI-INAF agreement no. 2022-28-HH.0 "PLATO Fase D". SNB and AFL acknowledge support from the INAF grant MASTODINT. CP thanks the Belgian Federal Science Policy Oﬃce (BELSPO) for the financial support in the framework of the PRODEX Program of the European Space Agency (ESA) under contract number 4000141194. S.M acknowledges support from the CNES GOLF-SOHO and PLATO grants at CEA/DAp.
LB and SM gratefully acknowledge support from the European Research Council (ERC) under the Horizon Europe programme (LB: Calcifer; Starting Grant agreement N$^\circ$101165631; SM: 4D-STAR; Synergy Grant agreement N$^\circ$101071505). While partially funded by the European Union, views and opinions expressed are, however, those of the authors only and do not necessarily reflect those of the European Union or the European Research Council. Neither the European Union nor the granting authority can be held responsible for them. The authors acknowledge G.~Buldgen, H.~Dhouib, and M.A.~Dupret for fruitful discussions. 
% \\
% \textit{Software:} 
% \texttt{mascaret} (this work),
% \texttt{numpy} \citep{harris2020array}, 
% \texttt{matplotlib} \citep{Hunter:2007}, 
% \texttt{scipy} \citep{2020SciPy-NMeth}, 
% \texttt{astropy} \citep{astropy:2022}, 
% \texttt{pandas} \citep{Pandas2020},
\end{acknowledgements}

\bibliographystyle{aa} 
\bibliography{biblio.bib} 

\appendix

\section{Model properties}

\begin{figure}[ht!]
    \centering
    \includegraphics[width=0.98\linewidth]{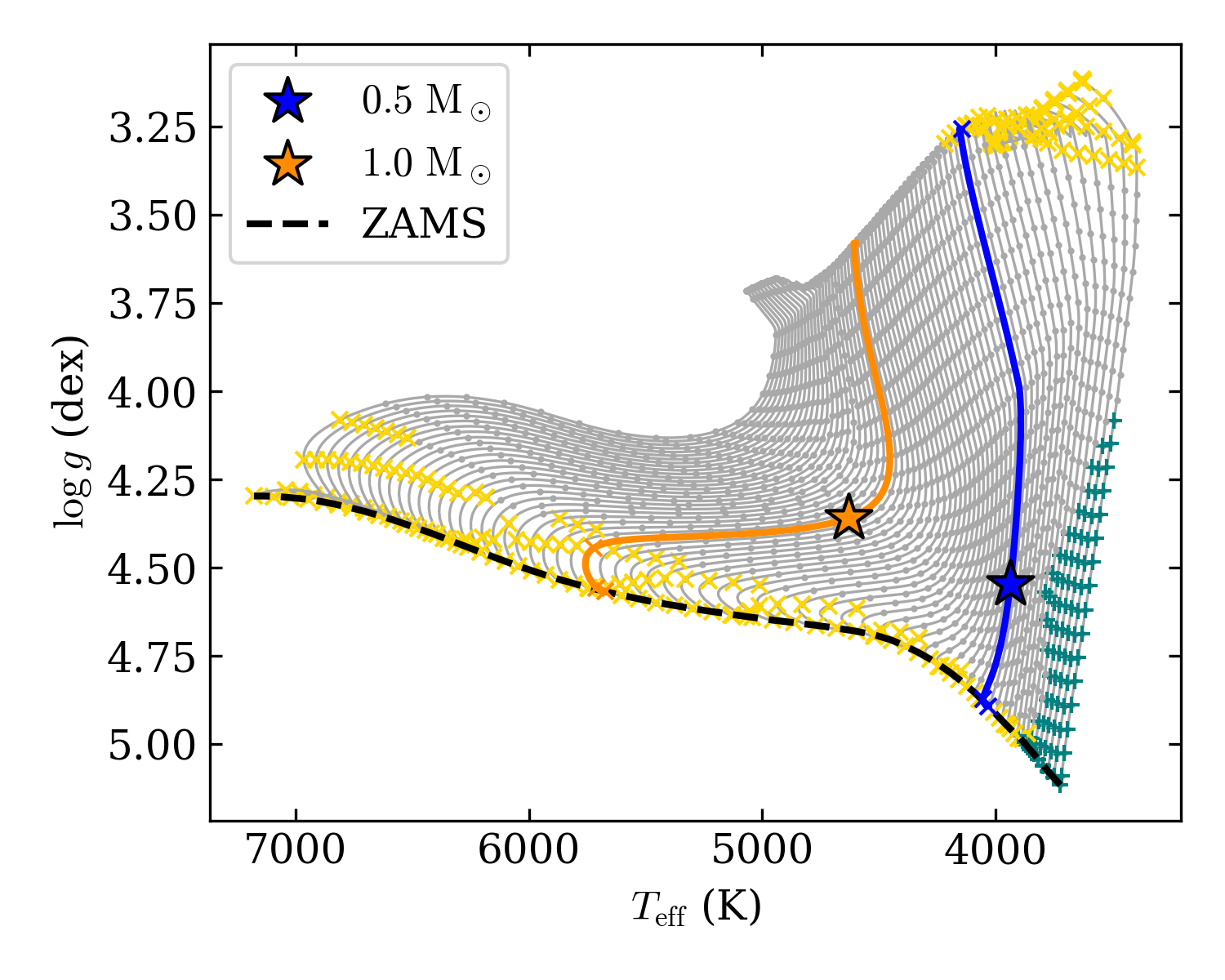}
    \caption{Evolutionary tracks for the stellar models (0.5~$\rm M_\odot$ in blue and 1~$\rm M_\odot$ in orange) considered in this work, from the end of the accreting phase to the ZAMS. The black thick lines shows the location of the ZAMS. The location on the track of the stellar models we consider in this work is highlighted by the star symbol. The other tracks of the \citet{Steindl2021} PMS grid ranging between $M_\star = 0.2 \; \mathrm{M}_\odot$ and $M_\star = 1.5 \; \mathrm{M}_\odot$ are shown in grey for comparison, with the structure model saved during the MESA run indicated by grey dots, yellow crosses, and teal plus signs. The crosses signal models which have a convective core while the plus signs correspond to stellar structure that are completely convective.}
    \label{fig:kiel_diagram}
\end{figure}

\begin{figure}[ht!]
    \centering
    \includegraphics[width=0.98\linewidth]{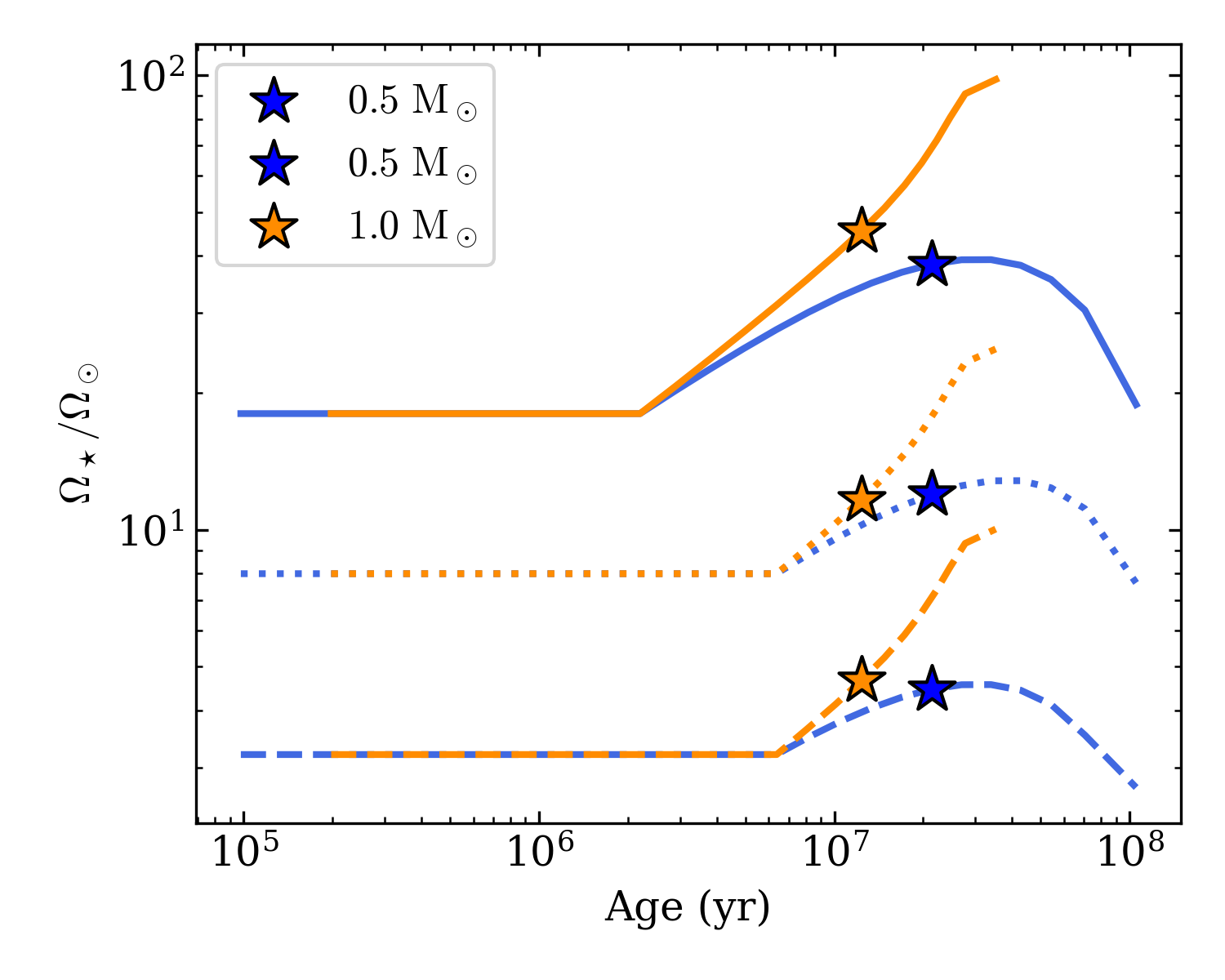}
    \caption{Rotational evolution computed for the 0.5 (blue) and 1~$\rm M_\odot$ (orange) models. The rotational track with slow initial condition is represented with a dashed line, {the track with intermediate initial condition with a dotted line}, and the track with fast initial condition with a thick line. The location of the stellar models we consider in this work is highlighted by the star symbol.}
    \label{fig:rotational_evolution}
\end{figure}

The properties of the stellar models we consider in this work, taken from the PMS stellar models with constant accretion computed by \citet{Steindl2021}, are summarised in Table~\ref{tab:properties}, while the evolutionary tracks computed with MESA and the rotational evolution history diagram are shown in Fig.~\ref{fig:kiel_diagram} and Fig.~~\ref{fig:rotational_evolution}, respectively. 
The accretion rate considered to reach the initial mass of each PMS model is $\num{5e-6} \; \mathrm{M}_\odot \mathrm{yr}^{-1}$.
In addition, Fig.~\ref{fig:kiel_diagram} shows the location of structure models exhibiting a convective core for the \citet{Steindl2021} models ranging between $M_\star = 0.2 \; \mathrm{M}_\odot$ and $M_\star = 1.5 \; \mathrm{M}_\odot$. We note that low-mass models have a convective core in the first epochs after the end of the accreting phase, while many models develop a small convective core just before entering the ZAMS. The lowest mass stars become completely convective before entering the ZAMS. 
{This justifies that the configuration we consider in this work (radiative interior surrounded by a convective envelope, with no convective core) is mostly adapted to PMS stars spun up by contraction, before they develop a convective core.}

\begin{table}[ht!]
    \caption{Summary of the PMS model properties.}
    \centering
    \begin{tabular}{ccc}
    \hline \hline
    Mass ($\rm M_\odot$) & 0.5 & 1 \\
    \hline
    Age (Myr) & 21.4 & 12.4 \\
    $M_\star$ ($\rm M_\odot$) & 0.5 & 1.0 \\
    $R_\star$ ($\rm R_\odot$) & 0.62 & 1.10 \\
    $\Omega_\star$ ($\rm \Omega_\odot$, slow case) & 4.5 & 4.7 \\
    $\Omega_\star$ ($\rm \Omega_\odot$, intermediate case) & 12.0 & 11.6 \\
    $\Omega_\star$ ($\rm \Omega_\odot$, fast case) & 38.4 & 45.4 \\
    \end{tabular}
    \label{tab:properties}
\end{table}

\section{Eigenfunction computation \label{appendix:eigenfunction computation}}

\subsection{System derivation \label{appendix:system_derivation}}

The problem is formulated considering the following set of equations in a differentially rotating star
\begin{align}
    D_t \bm{V} &= - \frac{1}{\hat{\rho}} \nabla \hat{p} - \nabla \hat{\Phi} \; , \\
    \diffp{\hat{\rho}}{t} &+ \nabla \cdot (\hat{\rho} \bm{V}) = 0 \; , \\
    \frac{1}{\Gamma_1} D_t \ln \hat{p} &- D_t \ln \hat{\rho}  = 0 \; , \\
    \nabla^2 \hat{\Phi} &= 4\pi G \hat{\rho} \; ,
\end{align}
where the equation are, in this order, the inviscid Euler equation of motion in an inertial reference frame, the equation of continuity, the energy equation in the adiabatic limit, and the Poisson's equation. $G$ is the gravitational constant and the Lagrangian derivative $D_t$ corresponds to $\partial / \partial t + \bm{V} \cdot \nabla$.
The scalar fields $\hat{p}$ (pressure),  $\hat{\rho}$ (density), and  $\hat{\Phi}$ (gravitational potential), are written in all generality and may depend on the time $t$ as well as any spatial coordinate. In spherical coordinates $(r, \theta, \varphi)$, associated with the unit vector basis $(\hat{\bm{e}}_r, \hat{\bm{e}}_\theta, \hat{\bm{e}}_\varphi)$, the velocity field $\bm{V}$ is 
\begin{equation}
    \bm{V} (r, \theta, \varphi, t) = r\sin\theta \Omega (r, \theta) \bm{\hat{e}_\varphi} + \bm{u} (r, \theta, \varphi, t) \; ,
\end{equation}
where $\bm{u}$ is the velocity field associated with the wave perturbation, related to the Lagrangian displacement, $\bm {\xi}$ through \citep{Unno1989}
\begin{equation}
    \bm{u} = \left(\diffp{}{t} + \Omega \diffp{}{\varphi}\right) \bm{\xi} - r \sin \theta (\bm{\xi} \cdot \nabla \Omega) \hat{\bm{e}}_{\varphi}  \; ,
\end{equation}

The linearised equations of motion, accounting for the Coriolis force are then, in the Cowling approximation \citep{Unno1989,Mathis2009}, having expanded the scalar fields $\hat{X}$ as the sum of an equilibrium quantity $X$ and a perturbation $\tilde{X}'$, and assuming that all the equilibrium quantities and the rotation profile $\Omega$ depend only on $r$ 
\begin{align}
    \left(\diffp{}{t} + \Omega \diffp{}{\varphi}\right) u_r - 2 \Omega \sin \theta u_\varphi &= - \frac{1}{\rho} \diffp{\tilde{p}'}{r} - \frac{\tilde{\rho}' g}{\rho}  \; , \\
    \left(\diffp{}{t} + \Omega \diffp{}{\varphi}\right) u_\theta - 2 \Omega \cos \theta u_\varphi  &= - \frac{1}{\rho r}\diffp{\tilde{p}'}{\theta} \; , \\
    \left(\diffp{}{t} + \Omega \diffp{}{\varphi}\right) u_\varphi + \zeta u_r \sin \theta + 2 \Omega \cos \theta u_\theta &= - \frac{1}{\rho r \sin \theta} \diffp{\tilde{p}'}{\varphi}  \; ,
\end{align}
where $g = \mathrm{d} \Phi / \mathrm{d}r$. The linearised continuity and energy equations are
\begin{align}
\begin{split}
    \left(\diffp{}{t} + \Omega \diffp{}{\varphi}\right) \rho' + u_r \diff{\rho}{r} &+ \rho \Bigg( \frac{1}{r^2} \diffp{r^2 u_r}{r} \\  &+ \frac{1}{r \sin \theta} \diffp{\sin \theta u_\theta}{\theta} + \frac{1}{r \sin \theta} \diffp{u_\varphi}{\varphi} \Bigg)  = 0 \; , 
\end{split} \\
    \left(\diffp{}{t} + \Omega \diffp{}{\varphi}\right) \left( \frac{\tilde{p}'}{\Gamma_1 p} - \frac{\tilde{\rho}'}{\rho} \right) &+ u_r \left( \frac{1}{\Gamma_1} \diff{\ln p}{r} - \diff{\ln \rho}{r} \right) = 0 \; .
\end{align}
Expanding the perturbed quantities $\tilde{X}'$ and the component of the displacement and velocity fields in Fourier series such that 
\begin{equation}
    \tilde{X}'(r, \theta, \varphi, t) = \sum_{k,m} X'(r) \exp [i (\omega_{\rm in} t + k \theta - m \varphi) ] \; ,
\end{equation}
where $k$ is the latitudinal wave number, and $\omega_{\rm in}$ is the inertial frequency, connected to the local frequency, $\omega$, through
\begin{equation}
    \omega (r) = \omega_{\rm in} - m \Omega (r) \; .
\end{equation}
In this convention, $m > 0$ corresponds to prograde modes and $m < 0$ to retrograde modes. Restricting the problem to the vicinity of the equatorial plane \citep[][Mathis et al. submitted]{Ando1985}, that is considering $\theta \approx \pi/2$, allows writing
\begin{align}
    \label{eq:motion_r}
    - \omega^2 \xi_r - 2 \Omega u_\varphi &= - \frac{1}{\rho} \diff{p'}{r} - \frac{\rho' g}{\rho}  \; , \\
    \label{eq:motion_theta}
    - \omega^2 \xi_\theta &= - \frac{ik}{\rho r} p'  \; , \\
    \label{eq:motion_phi}
    i \omega u_\varphi + i \omega \zeta \xi_r &= \frac{i m}{\rho r} p' \; , \\
    \label{eq:continuity}
    i \omega \rho' + i \omega \diff{\rho}{r} \xi_r  &+ \rho \Bigg( i \omega \Bigg[ \frac{2}{r} \xi_r  + \diff{\xi_r}{r} \Bigg] - \frac{k \omega}{r} \xi_\theta - \frac{i m}{r} u_\varphi \Bigg) = 0 \; , \\
    \label{eq:energy}
    \frac{p'}{\Gamma_1 p} &- \frac{\rho'}{\rho} + \frac{N^2}{g} \xi_r = 0 \; ,
\end{align}
where the Brunt-Väisälä frequency, $N$, is defined as 
\begin{equation}
    N^2 = g \left[ \frac{1}{\Gamma_1} \diff{\ln p}{r} - \diff{\ln \rho}{r} \right] \; .
\end{equation}
{This equatorial approximation is well suited to characterise the behaviour of gravito-inertial waves that are confined close to the equator \citep[e.g.][]{Townsend2003,Bekki2022b,Jain2023} without neglecting the non-traditional component of the Coriolis acceleration.}
Eq.~(\ref{eq:motion_theta}) and Eq.~(\ref{eq:motion_phi}) can be used to eliminate $\xi_\theta$ and $u_\varphi$ in Eq.~(\ref{eq:motion_r}) and (\ref{eq:continuity}) to yield 
\begin{align}
    \Bigg(- \omega^2 + 2\Omega\zeta \Bigg) \xi_r  - \frac{2\Omega m}{\rho r \omega} p' &= \; - \frac{1}{\rho} \diff{p'}{r} - \frac{\rho' g}{\rho} , \\
    \begin{split}
    i \omega \rho' + i \omega \diff{\rho}{r} \xi_r &+ \rho \Bigg( i \omega \Bigg[ \frac{2}{r} \xi_r  + \diff{\xi_r}{r} \Bigg] \\ &- \frac{i k^2}{\rho r^2 \omega} p' + \frac{i m \zeta}{r} \xi_r - \frac{i m^2}{\rho r^2 \omega} p' \Bigg) = 0 \; , 
    \end{split}
\end{align}
which can be rewritten 
\begin{align}
    \diff{p'}{r} &= \rho \Bigg( \omega^2 - 2\Omega\zeta \Bigg) \xi_r + \frac{2\Omega m}{\omega r} p' - \rho' g \; , \\
    \diff{\xi_r}{r} &= \left( - \diff{\ln \rho}{r} - \frac{2}{r} - \frac{m \zeta}{\omega r}\right) \xi_r + \frac{k^2 + m^2}{\rho r^2 \omega^2} p' - \frac{\rho'}{\rho} 
\end{align}
and finally, using Eq.~(\ref{eq:energy}) to express $\rho'$ in terms of $\xi_r$ and $p'$, and $g = - 1 / \rho \times (\mathrm{d}p / \mathrm{d}r)$ 
\begin{align}
    \diff{p'}{r} &= \rho \Bigg[\omega^2 - N^2 -2\Omega\zeta \Bigg] \xi_r + \Bigg[\frac{1}{\Gamma_1} \diff{\ln p}{r} + \frac{2\Omega m}{\omega r}  \Bigg] p' \; , \\
    \begin{split}
    \diff{\xi_r}{r} &= - \left[\frac{1}{r}\right(2 + \frac{m \zeta}{\omega}\left) + \frac{1}{\Gamma_1}\diff{\ln p}{r} \right] \xi_r \\ 
    &\qquad\qquad\qquad\qquad + \frac{1}{\rho c_s^2}\Bigg[\frac{1}{\omega^2}\frac{(k^2 + m^2) c_s^2}{r^2} - 1 \Bigg] p'  \; ,
    \end{split}
\end{align}
where, by making the identification ${\tilde{\ell}}({\tilde{\ell}}+1) = k^2 + m^2$ we obtain the system defined by Eq.~(\ref{eq:xi_r}) and (\ref{eq:p_prime}). In the case without rotation, the global horizontal number $\tilde{\ell}$ can be identified to the spherical degree $\ell$.

It should be noted that the behaviour of equatorial Rossby waves (also known as planetary waves) is not captured by such a model, because it does not include the topological $\beta$-effect (that is, the variation with latitude of the strength and direction of the Coriolis force) that is the direct cause for the propagation of the related perturbations. Nevertheless, it fully accounts for sphericity in the radial direction and the compressional $\beta$-effect responsible for the propagation of thermal Rossby waves. It is analogous to the millstone model presented by \citet{Jain2024} and the eigenmodes it allows computing in convective envelopes are among the mixed inertial modes they identified.

%{\color{red}Add a few considerations: this equatorial model captures the thermal Rossby waves (topological $\beta$-effect) but exclude equatorial Rossby waves which requires a latitudinal gradient in the Coriolis force.}

\subsection{Propagation conditions \label{appendix:propagation}}

We can combine Eq.~(\ref{eq:xi_r}) and (\ref{eq:p_prime}) to obtain an equation of the form 
\begin{equation}
    \diff[2]{\xi_r}{r} + A (r) \diff{\xi_r}{r} + B(r) \xi_r = 0 \;,    
\end{equation}
where, under the assumption of solid body rotation
\begin{align}
    A(r) &= \diff{}{r} \ln \left| \rho r^2 c_s^2 \left( \frac{S_\ell^2}{\omega^2} - 1 \right)^{-1} \right| \; \\
    \begin{split}
    B(r) &= \frac{1}{c_s^2} \left( \frac{S^2_{\tilde{\ell}}}{\omega^2} - 1 \right) \Bigg(N^2 + 4\Omega^2 - \omega^2 \Bigg) - \frac{4\Omega^2 m^2}{\omega^2 r^2} \\
    &- \frac{2 \Omega m}{\omega} \left[ \frac{3}{r^2} + \frac{1}{r} \left\{ \diff{}{r} \ln \left| \frac{1}{\rho c_s^2} \left( \frac{S_{\tilde{\ell}}^2}{\omega^2} - 1 \right)  \right| + \frac{2}{\Gamma_1} \diff{\ln p}{r} \right\} \right] \\
    &+ \diff{}{r} \left( \frac{1}{\Gamma_1} \diff{\ln p}{r} \right) - \frac{2}{r^2} - \left[ \frac{2}{r} + \frac{1}{\Gamma_1} \diff{\ln p}{r} \right] \diff{}{r} \ln \left| \frac{1}{\rho c_s^2} \left( \frac{S_{\tilde{\ell}}^2}{\omega^2} - 1 \right)  \right| \\
    &- \frac{2}{r}\frac{1}{\Gamma_1} \diff{\ln p}{r} - \frac{1}{\Gamma_1^2} \left(\diff{\ln p}{r} \right)^2 \; .
    \end{split}
\end{align}
%Refactor the $B$ terms in order to show the contribution evidenced by \citet{Jain2024} and in Stéphane's draft ?
We can then use the following change of variable
\begin{equation}
    \xi_r = C(r) \Psi (r) \; ,
\end{equation}
with
\begin{equation}
    C(r) = \exp \left( - \frac{1}{2} \int A(r) \mathrm{d}r \right) \; ,
\end{equation}
to write
\begin{equation}
    \diff[2]{\Psi}{r} + \hat{k}_r^2 \Psi = 0 \; ,
\end{equation}
with the radial wave vector, $\hat{k}_r^2$, given by
\begin{equation}
\label{eq:p2_schrodinger}
    \hat{k}_r^2 = B(r) + A(r) \diff{\ln C}{r} + \frac{1}{C} \diff[2]{C}{r} \; .
\end{equation}
The wave function $\Psi$ thus has an oscillating behaviour only in regions where $\hat{k}_r^2 > 0$ and is evanescent otherwise.
In the low-frequency $\omega^2 \ll S_{\tilde{\ell}}^2$ limit we have
\begin{align}
    \frac{1}{\rho c_s^2} \left( \frac{S_{\tilde{\ell}}^2}{\omega^2} - 1 \right) &= \frac{{\tilde{\ell}} ({\tilde{\ell}} +1)}{r^2 \omega^2} \; , \\
    C(r) &= \rho^{-1/2} r^{-2} \; ,
\end{align}
so that, similarly to \citet{Press1981}, we have $\Psi = \rho^{1/2} r^2 \xi_r$, and if we neglect the derivative of equilibrium quantities, we get
\begin{equation}
\begin{split}
    \hat{k}_r^2 \approx k_r^2 \equiv \frac{{\tilde{\ell}} ({\tilde{\ell}} + 1)}{r^2 \omega^2} \Bigg[ N_{\rm eff}^2 - \omega^2 \Bigg] 
   \; .
\end{split}
\end{equation}
where we defined
\begin{equation}
\label{eq:effective_stratification}
    N_{\rm eff}^2 \equiv N^2 + 4\Omega^2 r^2_{\tilde{\ell},m} \; ,
\end{equation}
with
\begin{equation}
    r_{\tilde{\ell},m} = \left( 1 - \frac{m^2}{{\tilde{\ell}} ({\tilde{\ell}} + 1)} \right)^{1/2} \; .
\end{equation}

In the limit $N^2 \gg 4 \Omega^2$, valid e.g. for the radiative interior, this yields the gravity wave propagation condition while the $4 \Omega^2 \gg N^2$ case corresponds to the pure inertial wave limit. This would occur for example in a convectively neutral medium where $N^2 = 0$. In a realistic stellar convective envelope, the superadiabaticity of the fluid has to be accounted for, and the wave can propagate only if  
\begin{equation}
\label{eq:propagation_condition_bulk}
    \omega^2 < N^2 + 4\Omega^2 r^2_{\tilde{\ell},m} \; ,
\end{equation}
where it should be kept in mind that $N^2 < 0$ in this specific case. Close to the surface, where the derivative of equilibrium quantities are not small with respect to the other terms in Eq.~(\ref{eq:p2_schrodinger}), the propagation condition can be extended to something of the form \citep{Hindman2023,Jain2023} 
\begin{equation}
    k_r^2 + \frac{2\Omega}{\omega \mathcal{H}} k_\varphi - h(r) > 0
\end{equation}
where we follow \citet{Jain2023} to define
\begin{equation}
    \frac{1}{\mathcal{H}} = \frac{1}{H_\rho} - \frac{2N^2}{g} \; ,
\end{equation}
where we introduced the density scale height $H_\rho = - \mathrm{d}r / \mathrm{d} \ln \rho$, and $k_\varphi = m/r$ is the azimuthal wave number, while the $h$ term depends on the derivative of the equilibrium quantities. Under the Lantz-Braginsky-Roberts \citep{Lantz1992,Braginsky1995} anelastic approximation, it can be shown that $h$ reduces to the acoustic cutoff wavenumber, $k_c^2$,
\begin{equation}
    h \approx k_c^2 \equiv \frac{1}{4 H_\rho} \left(  1 - 2 \diff{H_\rho}{r} \right)
\end{equation}

\subsection{Inertial dips \label{appendix:inertial_dips}}

In this section, we adapt the procedure outlined by \citet[][hereafter TT22]{Tokuno2022} to derive an analytic approximation for the profile of the inertial dips observed in Fig.~\ref{fig:modes_both_profiles}. 

\subsubsection{Formulation in the radiative interior}

We start by noting that, given the form of $k_r$, for modes with $\omega^2 < 4 \Omega^2 r^2_{\tilde{\ell},m}$, $k_r$ does not strictly cancel out in the vicinity of the convective-radiative interface and close to the stellar centre. Nevertheless, in the vicinity of these two regions, as long as $\Omega$ is not too large with respect to $\omega$, we have $\omega \sim N_{\rm eff}$ and $k_r$ small enough in order to apply the asymptotic treatment that will allow us to obtain an analytic expression for $\Psi$. We consider that these pseudo-turning points are located at $r_a$ and $r_b$, located close to the stellar centre and to the convective-radiative interface, respectively. Assuming that we have have the following differential equation for $\Psi$ 
\begin{equation}
    \diff[2]{\Psi}{r} + k_r^2 \Psi \approx 0 \; ,
\end{equation}
and that it is valid in the vicinity of both $r_a$ and $r_b$\footnote{Unlike the case discussed in \citetalias{Tokuno2022}, the two (pseudo-)turning points we consider here have the same nature and correspond to $\omega \sim N_{\rm eff}$.}, we therefore have \citep{Unno1989}
\begin{equation}
    \Psi_1 (r) \approx \frac{1}{|k_r|^{1/2}} \left( \frac{3}{2} \left| \int_{r_a}^r |k_r| \mathrm{d}r \right| \right)^{1/6} \\ \times [a \mathrm{Ai} (z_1) + b \mathrm{Bi} (z_1)] \; ,   
\end{equation} 
for $r \ll r_b$ and
\begin{equation}
    \Psi_2  (r) \approx \frac{1}{|k_r|^{1/2}} \left( \frac{3}{2} \left| \int_r^{r_b} |k_r| \mathrm{d}r \right| \right)^{1/6} \\ \times [c \mathrm{Ai} (z_2) + d \mathrm{Bi} (z_2)] \; , 
\end{equation} 
for $r_a \ll r$, with $a$, $b$, $c$, $d$ are constants to determine from the boundary conditions. $\rm Ai$ and $\rm Bi$ are the Airy functions of first and second kind while $z_1$ and $z_2$ are given by. 
\begin{align}
    z_1 &\equiv \mathrm{sgn} (k_r^2) \left( \frac{3}{2} \left| \int_{r_a}^r |k_r| \mathrm{d}r \right| \right)^{2/3} \; , \\ 
    z_2 &\equiv \mathrm{sgn} (k_r^2) \left( \frac{3}{2} \left| \int_r^{r_b} |k_r| \mathrm{d}r \right| \right)^{2/3} \; ,     
\end{align}
First, we have $b = 0$ to ensure that the functions decay exponentially below $r_a$, close to the centre of the star. Then, by matching $\Psi_1$ to $\Psi_2$ at $r_a \ll r \ll r_b$, it comes that
\begin{align}
    c &= a \cos \left ( \int_{r_a}^{r_b} k_r \mathrm{d}r - \frac{\pi}{2}\right) = a \sin \left( \frac{\pi^2 s}{\Omega \Delta P_0} \right)  \\
    d &= - a \sin \left ( \int_{r_a}^{r_b} k_r \mathrm{d}r - \frac{\pi}{2}\right) = a \cos \left( \frac{\pi^2 s}{\Omega \Delta P_0} \right)
\end{align}
where, introducing for convenience the spin parameter $s = 2\Omega/\omega$, we have defined \citep{Tassoul1980}
\begin{equation}
\label{eq:tassoul_delta_p}
    \Delta P_0 \equiv \frac{2\pi^2}{\sqrt{{\tilde{\ell}} ({\tilde{\ell}}+1)}} \left( \int_{r_a}^{r_b} \frac{N^2_{\rm eff}}{r} \mathrm{d}r \right)^{-1} \; ,
\end{equation}
and we have computed the $\int k_r$ term assuming that $\omega \ll N$. Next, we need to express $\Psi$ and $\mathrm{d}\Psi/\mathrm{d}r$ at $r_b$. A Taylor expansion around $r_b$ yields
\begin{equation}
\label{eq:taylor_development_kr}
    k_r^2 \approx \left[ - \frac{{\tilde{\ell}} ({\tilde{\ell}}+1)}{r^2 \omega^2} \diff{N^2}{r} \right]_{r = r_b} (r_b - r) = \frac{{\tilde{\ell}}({\tilde{\ell}}+1) s^2}{\varepsilon^3 r_b^3} (r_b - r)
\end{equation}
where, under solid-body rotation, $\mathrm{d}N_{\rm eff}^2 / \mathrm{d}r = \mathrm{d}N^2 / \mathrm{d}r$, and we followed \citetalias{Tokuno2022} to define
\begin{equation}
\label{eq:epsilon}
    \varepsilon \equiv \left( - \frac{r_b}{4 \Omega^2} \diff{N^2}{r} \Bigg|_{r = r_b} \right)^{-1/3} \; .
\end{equation}
We note that for $r$ close to $r_b$ with $r < r_b$, $\mathrm{d}N^2 / \mathrm{d}r < 0$. We have
\begin{align}
    &\lim_{r\rightarrow r_b} \frac{1}{|k_r|^{1/2}} \left( \frac{3}{2} \left| \int_{r_a}^r |k_r| \mathrm{d}r \right| \right)^{1/6} = \frac{\varepsilon^{1/2} r_b^{1/2}}{s^{1/3} [{\tilde{\ell}} ({\tilde{\ell}}+1)]^{1/6}} \; , \\
    &\lim_{r\rightarrow r_b} \diff{}{r} \left[ \frac{1}{|k_r|^{1/2}} \left( \frac{3}{2} \left| \int_{r_a}^r |k_r| \mathrm{d}r \right| \right)^{1/6} \right] = 0 \; , \\
    &\lim_{r\rightarrow r_b} \diff{z_2}{r} = \frac{s^{2/3} [{\tilde{\ell}}({\tilde{\ell}}+1)]^{1/3}}{\varepsilon r_b}
\end{align}
so that
\begin{equation}
\begin{split}
    \Psi (r_b) = \frac{2}{3^{2/3}\Gamma(2/3)} \frac{a r_b^{1/2} \varepsilon^{1/2}}{s^{1/3}[{\tilde{\ell}}({\tilde{\ell}}+1)]^{1/6}} \cos \left( \frac{\pi^2 s}{\Omega \Delta P_0} - \frac{\pi}{6} \right) \; ,
\end{split}
\end{equation}
and
\begin{equation}
    \diff{\Psi}{r} \Bigg|_{r=r_b} = \frac{2}{3^{1/3}\Gamma(1/3)}\frac{a s^{1/3} [{\tilde{\ell}}({\tilde{\ell}}+1)]^{1/6}}{r_b^{1/2} \varepsilon^{1/2}} \cos \left(  \frac{\pi^2 s}{\Omega \Delta P_0} - \frac{5\pi}{6} \right) \; ,
\end{equation}
where $\Gamma$ denotes the Gamma function. In what follows, we assume that this approximation for $\Psi$ is valid at $r_b = r_{\rm CZ}$.

\subsubsection{Formulation in the convective envelope}

In order to find an expression for the wave behaviour in the convective envelope, we look for a solution in the low-frequency limit $\omega^2 \ll S_{\tilde{\ell}}^2$, assuming that the wave is trapped between two rigid boundaries at the convective-radiative interface and at the surface, so that 
\begin{equation}
    \xi_r (r_{\rm CZ}) = \xi_r (R_\star) = 0 \implies \Psi (r_{\rm CZ}) = \Psi (R_\star) = 0  \; ,
\end{equation}
We also simplify the problem by assuming $N^2 = 0$ everywhere in the convective envelope. This leads to 
\begin{equation}
    k^2_r = \frac{{\tilde{\ell}}({\tilde{\ell}}+1)}{r^2 \omega^2} \left[ 4\Omega^2 r^2_{\hat{\ell},m} - \omega^2 \right] \; ,
\end{equation}
where $k^2_r > 0$ everywhere when $\omega < 2 \Omega r_{\tilde{\ell},m}$. We therefore write $\Psi$ as
\begin{equation}
    \Psi (r) = A k_r^{-1/2} \cos \left( \int_r^{R_\star} k_r \mathrm{d}r + \phi \right) \; ,
\end{equation}
where $A$ is an arbitrary constant. The condition $\Psi (R_\star) = 0$ yields $\phi = \pi/2$
and $\Psi (r_{\rm CZ}) = 0$ then provides the quantification condition as
\begin{equation}
    \sin \left( \int_{r_{\rm CZ}}^{R_\star} k_r \mathrm{d}r \right) = 0 \implies \int_{r_{\rm CZ}}^{R_\star} k_r \mathrm{d}r = n_* \pi \; ,
\end{equation}
where $n_*$ is the radial order of the envelope inertial mode. Given the $N^2 = 0$ assumption, we have an analytical solution for $\int k_r$
\begin{equation}
    \int_r^{R_\star} k_r \mathrm{d}r = \frac{\sqrt{{\tilde{\ell}}({\tilde{\ell}}+1)}}{\omega} \left[ 4\Omega^2 r^2_{\hat{\ell},m} - \omega^2 \right]^{1/2} \ln \left( \frac{R_\star}{r} \right) \; ,
\end{equation}
so that 
\begin{equation}
\begin{split}
    \Psi (r) = - &A \frac{r^{1/2}}{[{\tilde{\ell}}({\tilde{\ell}}+1)]^{1/4} s^{1/2}} \left[ r^2_{\hat{\ell},m} - \frac{1}{s^2} \right]^{-1/4} \\
    &\times \sin \left( \sqrt{{\tilde{\ell}}({\tilde{\ell}}+1)} s \left[ r^2_{\hat{\ell},m} - \frac{1}{s^2} \right]^{1/2} \ln \left( \frac{R_\star}{r} \right) \right) \; .
\end{split}
\end{equation}
The latter is null at $r_{\rm CZ}$ for the pure inertial wave trapped in the convective envelope but not when we will be looking for a continuous solution at the convective-radiative interface. We have also
\begin{equation}
\begin{split}
    \diff{\Psi}{r}\Bigg|_{r=r_{\rm CZ}} = A r_{\rm CZ}^{-1/2} [{\tilde{\ell}}({\tilde{\ell}}+1)]^{1/4} s^{1/2} \left[ r^2_{\hat{\ell},m} - \frac{1}{s^2} \right]^{1/4} \\
    \times \cos \left( \sqrt{{\tilde{\ell}}({\tilde{\ell}}+1)} s \left[ r^2_{\hat{\ell},m} - \frac{1}{s^2} \right]^{1/2} \ln \left( \frac{R_\star}{r_{\rm CZ}} \right) \right) \; .
\end{split}
\end{equation}

\subsubsection{Continuity condition}

\begin{figure}[ht!]
    \centering
    \includegraphics[width=0.99\linewidth]{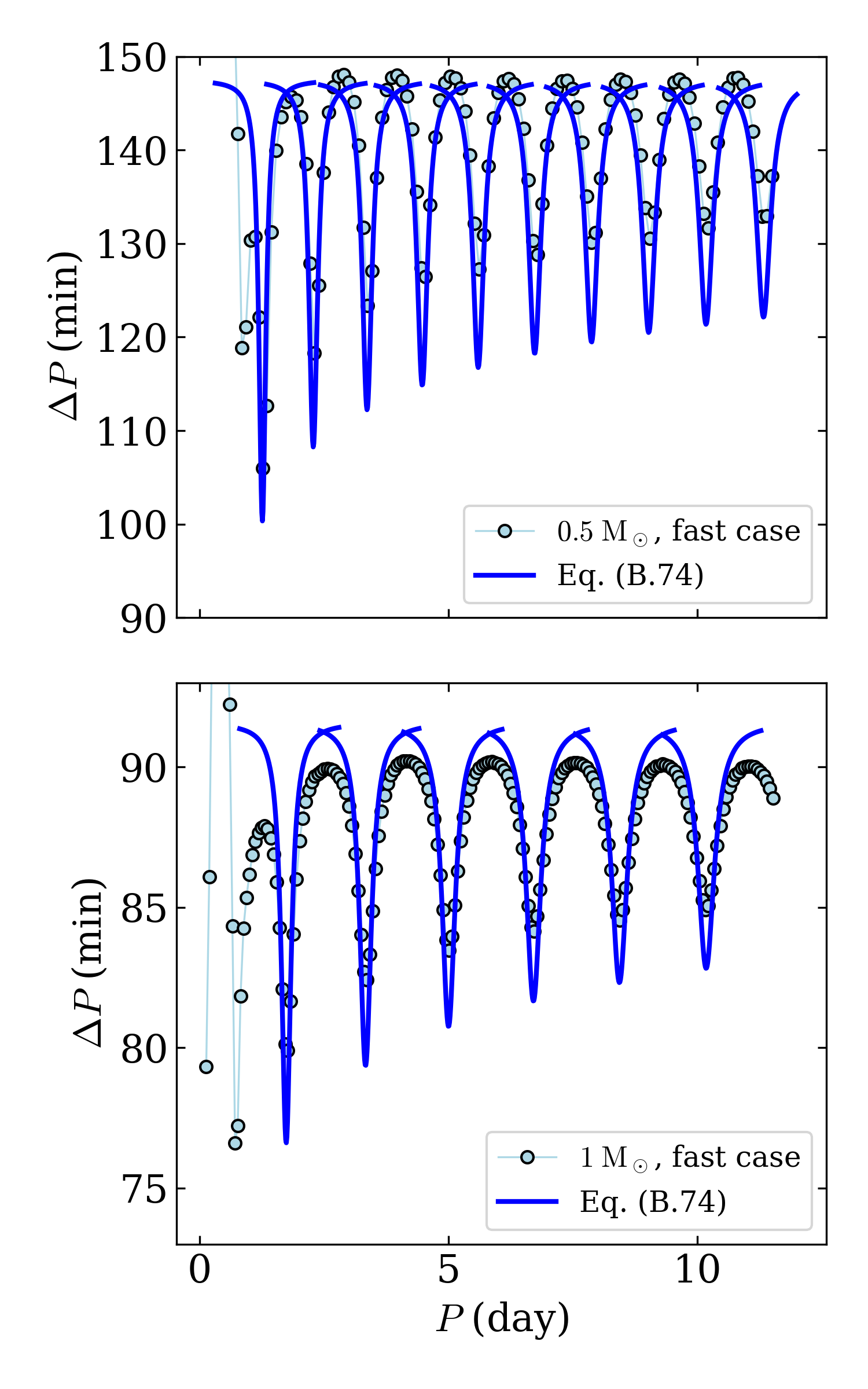}
    \caption{\textit{Top:} $\Delta P$ vs $P$ diagram for the 0.5~$\rm M_\odot$ model in the fast rotating case. The asymptotic Lorentzian profile defined in Eq.~(\ref{eq:lorentzian_profile}) are computed in the vicinity of the eigenfrequency of each inertial envelope mode.
    \textit{Bottom:} Same as top panel for the 1~$\rm M_\odot$ model.}
    \label{fig:dips_both_models}
\end{figure}

The Lagrangian pressure perturbation $\delta p$ and the displacement $\xi_r$ must be continuous at $r_{\rm CZ}$. Given that the equilibrium quantity $\rho$ and $p$ are continuous, it is enough to ensure that $\Psi$ and $\mathrm{d}\Psi / \mathrm{d}r$ are too, so that
\begin{align}
    \label{eq:continuity_psi}
    \Psi (r_{\rm CZ,-}) = \Psi (r_{\rm CZ,+}) \; , \\
    \label{eq:continuity_dpsi_dr}
    \diff{\Psi}{r} \Bigg|_{r = r_{\rm CZ,-}} = \diff{\Psi}{r} \Bigg|_{r = r_{\rm CZ,+}} \; .
\end{align}
Substituting the terms by their analytical expression and dividing Eq.~(\ref{eq:continuity_psi}) by (\ref{eq:continuity_dpsi_dr}), we get
\begin{equation}
\begin{split}
     \varepsilon &\left[ \tan \left( \frac{\pi^2 s}{\Omega \Delta P_0} - \frac{\pi}{6} \right) - \frac{1}{\sqrt{3}} \right]^{-1} = \\
     &\left( \frac{3^{1/3}\Gamma(2/3)}{\Gamma(1/3)} \right) \frac{1}{ [{\tilde{\ell}}({\tilde{\ell}}+1)]^{1/6} s^{1/3}} \left[ r^2_{\hat{\ell},m} - \frac{1}{s^2} \right]^{-1/2} \\
     &\times \tan \left( \sqrt{{\tilde{\ell}}({\tilde{\ell}}+1)} s \left[ r^2_{\hat{\ell},m} - \frac{1}{s^2} \right]^{1/2} \ln \left( \frac{R_\star}{r_{\rm CZ}} \right) \right) \; .
\end{split}
\end{equation}
It is then possible to expand the right-hand term in the vicinity of the point where it vanishes, located at $s_*$, which is defined as the spin parameter of the isolated envelope inertial mode. We obtain
\begin{equation}
\label{eq:tangent_relation}
    \tan \left( \frac{\pi^2 s}{\Omega \Delta P_0} - \frac{\pi}{6} \right) = \frac{\varepsilon}{ (s - s_*) V } + \frac{1}{\sqrt{3}} \; ,
\end{equation}
where the term $V$ is given by
\begin{equation}
\begin{split}
    V = &\left( \frac{3^{1/3}\Gamma(2/3)}{\Gamma(1/3)} \right) \ln \left( \frac{R_\star}{r_{\rm CZ}} \right) \\
    &\times \frac{[{\tilde{\ell}} ({\tilde{\ell}}+1)]^{5/6}}{s_*^{1/3}} \left[ 1 + \frac{1}{s_*^2} \left(r^2_{\hat{\ell},m} - \frac{1}{s_*^2} \right)^{-1/4}  \right] \; .
\end{split}
\end{equation}
We now consider two neighbouring modes with spin parameters $s_1$ and $s_2$, and periods $P_1$ and $P_2$ such that $s_2>s_1$, $P_2>P_1$, where we recall that $s = \Omega P / \pi$. We also define $\overline{s} = (s_1 + s_2) / 2$. We have \citepalias{Tokuno2022}
\begin{equation}
    \frac{\pi^2 (s_2 - s_1)}{\Omega \Delta P_0} - \pi = y \ll 1 \,
\end{equation}
If we follow \citet{Barrault2025} and note $\overline{S} = \pi^2 \overline{s}/{\Omega \Delta P_0} - \pi/6$, we obtain the following relation from Taylor expansion
\begin{align}
\label{eq:taylor_expansion_S_1}
    \tan (\overline{S} + y) = \tan (\overline{S}) + y (1 + \tan^2 \overline{S})  \; , \\
\label{eq:taylor_expansion_S_2}
    \tan (\overline{S} - y) = \tan (\overline{S}) - y (1 + \tan^2 \overline{S})  \; ,.     
\end{align}
By subtraction of Eq.~(\ref{eq:taylor_expansion_S_1}) and (\ref{eq:taylor_expansion_S_2}), considering the relation derived in Eq.~(\ref{eq:tangent_relation}), this leads to
\begin{equation}
\begin{split}
    2 &\left[ 1 + \left( \frac{\varepsilon / V}{(\overline{s} - s_*) } + \frac{1}{\sqrt{3}} \right)^2  \right] \left[ \frac{\pi^2 (s_2 - s_1)}{\Omega \Delta P_0} - \pi \right] = \\
    &\frac{\varepsilon}{V} \left[ \frac{1}{s_2 + (s_2 - s_1) / 2 - s_*} - \frac{1}{s_1 + (s_1 - s_2) / 2 - s_*} \right] \; .
\end{split}
\end{equation}
Assuming that $(s_1 - s_*)(s_2 - s_*) \approx (\overline{s} - s_*)^2$, we recover
\begin{equation}
    \left[ 1 + \left( \frac{\varepsilon / V}{(\overline{s} - s_*) } + \frac{1}{\sqrt{3}} \right)^2  \right] \left[ \frac{\pi^2 (s_2 - s_1)}{\Omega \Delta P_0} - \pi \right] = - \frac{\varepsilon}{V} \frac{s_2 - s_1}{(\overline{s} - s_*)^2} \; ,
\end{equation}
which is Eq.~(63) from \citetalias{Tokuno2022}. The main difference is that the function $V$, which contains the properties of the envelope inertial mode, is different. Defining $\Delta P = P_2 - P_1$, $P = \Omega \overline{s} / \pi$, and $P_* = \Omega s_* / \pi$, we finally obtain our asymptotic relation for the inertial dip profile
\begin{equation}
\label{eq:lorentzian_profile}
    \Delta P = \Delta P_0 \left( 1 - \frac{\Delta P_0 \sigma / \pi}{(P - P_* + \sigma / \sqrt{3})^2 + \sigma^2 + \Delta P_0 \sigma / \pi} \right) \; ,
\end{equation}
where
\begin{equation}
\label{eq:sigma}
    \sigma \equiv \frac{3 \pi \varepsilon}{4 \Omega V} \; .
\end{equation}
As in \citetalias{Tokuno2022} and \citet{Barrault2025}, this means that the inertial dip can be described by the Lorentzian function defined in Eq.~(\ref{eq:lorentzian_profile}). In Fig.~\ref{fig:dips_both_models}, we show the Lorentzian profile computed in the vicinity of each inertial envelope eigenmode obtained in the fast rotating case, for both the 0.5~$\rm M_\odot$ and the 1~$\rm M_\odot$ models. The agreement between the asymptotic analytical solution and the numerical solutions is very good and correctly describes the shape of the inertial dip. 

\subsubsection{Frequencies in the inertial frame \label{sec:inertial_frame_frequencies}}

\begin{figure}[ht!]
    \centering
    \includegraphics[width=0.99\linewidth]{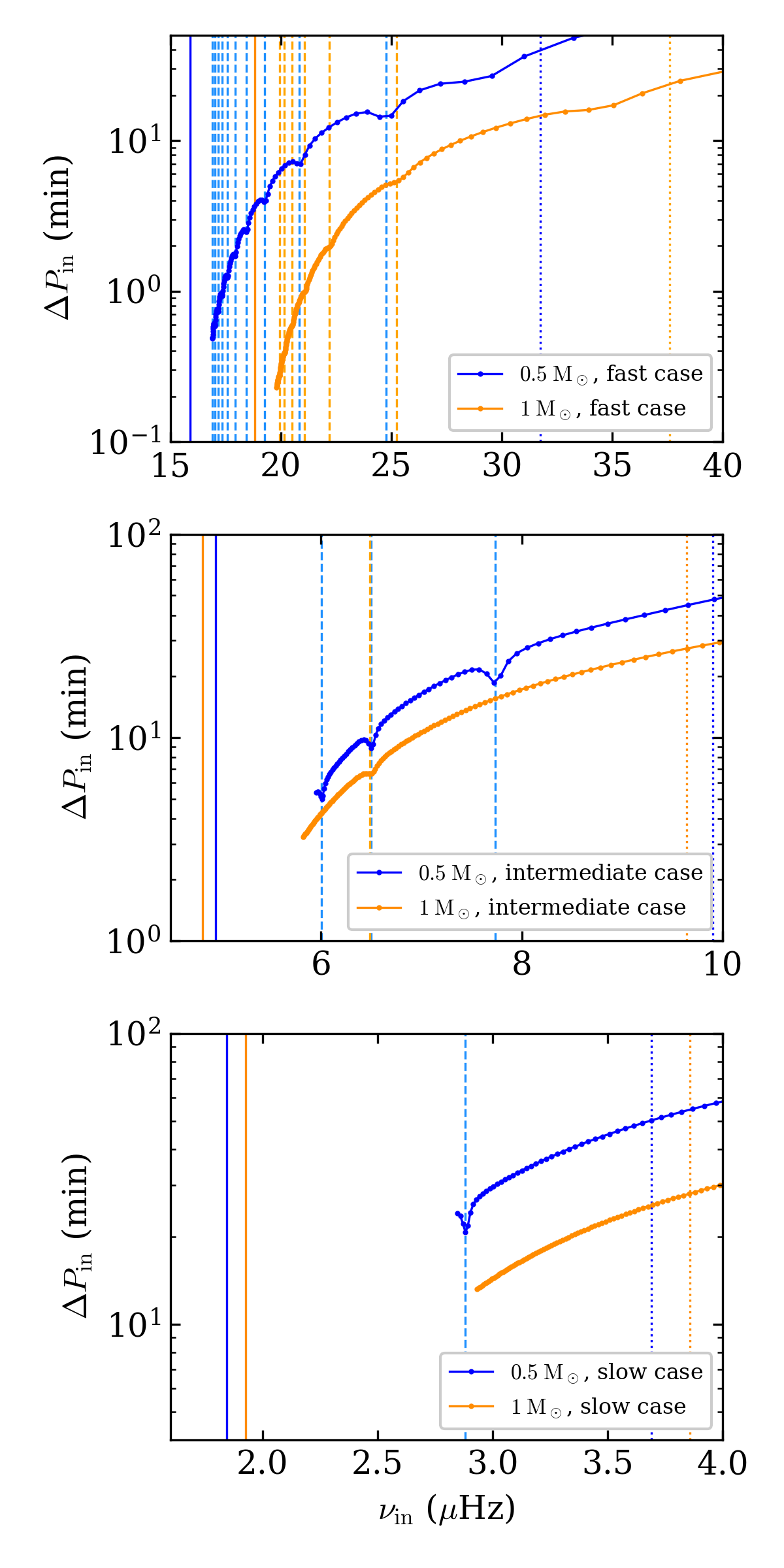}
    \caption{\textit{Top:} $\Delta P_{\rm in}$ vs $\nu_{\rm in}$ diagram for the fast rotation case. The modes computed for the 0.5~$\rm M_\odot$ and 1~$\rm M_\odot$ models are represented in orange and blue, respectively, each dot corresponding to a mode. The eigenfrequencies of the envelope inertial modes are highlighted with the vertical dashed lines, in orange and light blue, respectively. For each model, the first and second harmonics of the rotation period are shown by the thick and dotted lines, respectively.
    \textit{Middle:} Same as top panel for the intermediate rotation case.
    \textit{Bottom:} Same as top panel for the slow rotation case.}
    \label{fig:delta_p_inertial_frame_frequency_xaxis}
\end{figure}

The observed frequencies will be the ones from the inertial frame, $\omega_{\rm in} = \omega + m\Omega = 2 \pi \nu_{\rm in}$, and we can define the corresponding inertial period spacing, $\Delta P_{\rm in}$, as the period difference between two consecutive periods in the inertial frame. In Fig~\ref{fig:delta_p_inertial_frame_frequency_xaxis}, we show the $\Delta P_{\rm in}$ vs $\nu_{\rm in}$ diagram for both the 0.5~$\rm M_\odot$ and the 1~$\rm M_\odot$ models, in the {three rotation regimes, slow, intermediate, and fast}.  First, we can note that the profile of the inertial dips is more pronounced in the case of the 0.5~$\rm M_\odot$ model. {The range of inertial frequency where the dips has to be searched also depends on the rotation regime, with the dips occurring at higher frequency for the fast rotating case (between 15 and 30~$\mu$Hz), than for the intermediate (below 8~$\mu$Hz) and slow (below 3~$\mu$Hz) cases. While the low-frequency location of the first dip computed in the 0.5~$\rm M_\odot$ slow case would make it challenging to characterise, the frequency range of the dip occurrence in the intermediate case should be much more easily accessible to observations granted that the temporal baseline is sufficient and the instrument is stable enough.} 

{It can finally be seen that the modes where the inertial dips occur are located between the first and second harmonics of rotation. Indeed,} as expected from the relation between the inertial and co-rotation frequencies, for $m=1$ modes, $\omega_{\rm in}$ will asymptotically approach $\Omega$ as $\omega$ decreases (that is, as the mode absolute radial order increases). {Around these harmonics, large amplitude modulations from surface active regions can be expected. Nevertheless, if detectable, the signature from the gravito-inertial modes should be unambiguously distinguishable from the rotation peaks. Indeed, they are not located at the exact same frequency as the rotation peaks, and, more importantly, they will be asymmetrically distributed with respect to them, allowing for the characterisation of their regular pattern.}

\subsection{Numerical approach \label{appendix:numerical_approach}}

\subsubsection{Adimensioned equations}

Oscillation equations are more conveniently numerically solved under an adimensioned formulation. By introducing the adimensioned coordinate $x = r / R$ and the new variables $y_1$ and $y_2$ as defined by \citet{Townsend2013}
\begin{align}
    y_1 &= \frac{\xi_r}{r} x^{2-{\tilde{\ell}}} \; , \\
    y_2 &= \frac{p'}{\rho gr} x^{2-{\tilde{\ell}}} \; ,
\end{align}
where $g$ is the gravitational field, Eq.~(\ref{eq:xi_r}) and (\ref{eq:p_prime}) can be adimensioned, which yields
\begin{align}
\label{eq:adimensioned_y1}
    x \diff{y_1}{x} &= \left[ V_g - {\tilde{\ell}} - 1 - \frac{m\overline{\zeta}}{\overline{\omega}} \right] y_1
    + \left[\frac{{\tilde{\ell}}({\tilde{\ell}}+1)}{C \overline{\omega}^2} - V_g \right] y_2
    \; , \\
\label{eq:adimensioned_y2}
    x \diff{y_2}{x} &= \left [ C \bigg( \overline{\omega}^2 - 2\overline{\Omega}\overline{\zeta} \bigg) - A^{*} \right] y_1
    + \left [ A^{*} + m \frac{2\overline{\Omega}}{\overline{\omega}} - {\tilde{\ell}} + 3 - U \right] y_2 
    \; ,
\end{align}
where \citep[see][]{Dzembowski1971}
\begin{align}
    C &= x^3 \frac{M}{M_r} \;, \\
    V_g &= - \frac{1}{\Gamma_1} \diff{\ln p}{\ln r} = \frac{1}{\Gamma_1} \frac{\rho g r}{p} \;, \\ 
    U &= \frac{4\pi\rho r^3}{M_r} \;, \\
    A^{*} &= \frac{N^2 r}{g} \; ,
\end{align}
with $M_r$ the mass contained within the sphere of radius $r$. The adimensioned frequencies $\overline{\omega}$, $\overline{\Omega}$, and $\overline{\zeta}$ are defined as $\overline{\omega} (x) = \omega (x) / \Omega_{\rm c}$, $\overline{\Omega} (x) = \Omega (x) / \Omega_{\rm c}$, and $\overline{\zeta} (x) = \zeta (x) / \Omega_{\rm c}$. 

\subsubsection{Boundary conditions}

The boundary conditions we use are the following. First, we impose that the Lagragian pressure perturbation, $\delta p$, vanishes at the surface of the star ($x=1$)
\begin{equation}
    \delta p = p' + \xi_r \diff{p}{r} = 0 \; ,
\end{equation}
which translates to 
\begin{equation}
    y_1 (1) - y_2 (1) = 0 \; .
\end{equation}
Next, in order to regularise the problem at the centre of the star, we choose $\Omega$ to be a smooth profile that vanishes at $x = 0$
\begin{equation}
   \Omega (x) = (1 - \exp [-x/x_0]) \Omega_\star \; ,
\end{equation}
where we choose $x_0 = \num{1e-4}$. This way, given that for $x \rightarrow 0$, $A^{*} \sim x^3$, $V_g \sim x^2$, and $U \sim 3$, the regularisation condition at the centre is 
\begin{equation}
    C \overline{\omega}^2 (0) y_1 (0) - {\tilde{\ell}} y_2 (0) = 0\;.
\end{equation}
Finally, we choose $\overline{\omega} (1)$ as the reference frequency to solve the system.

\end{document}